\documentclass[12pt,a4paper]{article}
\usepackage[margin=28mm]{geometry}

\usepackage[T1]{fontenc}
\usepackage[T1,hyphens]{url}
\usepackage{amsfonts, amsmath, amssymb}  
\usepackage{tabularray}
\usepackage{nicematrix}
\NiceMatrixOptions{light-syntax,renew-dots}
\UseTblrLibrary{amsmath}
\usepackage{float}
\usepackage{parskip}
\usepackage{graphicx}
\usepackage{soul}
\usepackage{pifont} 
\usepackage{xspace}
\usepackage{todonotes}

\newcommand{\pobs}[1]{#1}
\newcommand{\obs}[1]{\mathsf{\pobs{#1}}}

\newcommand{\Ux}{\mathsf{U}_x}
\newcommand{\Uxt}{\widetilde{\mathsf{U}}_x}
\newcommand{\Cxt}{\widetilde{\mathsf{C}}_x}
\newcommand{\Cxtm}{\widetilde{\mathsf{C}}'_x}
\newcommand{\dg}{^{\dagger}}
\newcommand{\Hmin}{H_{\min}}

\newtheorem{theorem}{Theorem}
\newtheorem{remark}{Remark}

\usepackage{physics}

\usepackage{times}
\usepackage{xcolor,colortbl}
\usepackage{caption}
\usepackage{multicol}
\usepackage{graphicx}
\usepackage{url}
\usepackage{hyperref}
\usepackage{orcidlink}

\usepackage[english]{babel}
\usepackage{hyphenat}

\newcommand{\figpdf}[3]{%
    \begin{figure}[#3]
    \centering
        \includegraphics{Figures/#1.pdf}
    \caption{\small{#2}}
    \label{#1}
    \end{figure}
}

\def\({\left(}
\def\){\right)}

\newcommand{\C}{\mathbb{C}}
\newcommand{\hilbert}{\mathcal{H}}

\newcommand{\orcid}[1]{\href{https://orcid.org/#1}{\textcolor[HTML]{A6CE39}{\aiOrcid}}}

\title{Testing the 3D QRNG by Undoing}

\author{J. M. Ag\"{u}ero Trejo\footnote{School of Computer Science, University of Auckland, New Zealand. Email: \href{mailto:jagu688@aucklanduni.ac.nz}{jagu688@aucklanduni.ac.nz}, \href{mailto:manuel.aguero15@gmail.com}{manuel.aguero15@gmail.com}.}\ \orcidlink{0000-0001-9631-0326}, 
Cristian S. Calude\footnote{School of Computer Science, University of Auckland, New Zealand. Email: \href{mailto:c.calude@auckland.ac.nz}{c.calude@auckland.ac.nz}, \href{mailto:cscalude@gmail.com}{cscalude@gmail.com}.}\ \orcidlink{0000-0002-8711-6799}, 
O. C. Stoica\footnote{Department of Theoretical Physics, NIPNE--HH, Bucharest, Romania. 	Email: \href{mailto:cristi.stoica@theory.nipne.ro}{cristi.stoica@theory.nipne.ro}, \href{mailto:holotronix@gmail.com}{holotronix@gmail.com.} 
Corresponding author.
} \ \orcidlink{0000-0002-2765-1562}}

\date{\today}

\begin{document}

\maketitle

\begin{abstract}
We propose a method to test whether a photonic 3D QRNG operates according to the underlying theory by realizing the proposed measurement with the correct output probabilities and value-indefiniteness conditions, thereby generating strongly incomputable and maximally unpredictable sequences of random digits. The test relies on undoing the unitary evolution realized by the 3D  QRNG. The test verifies the unitarity of the realized transformation, quantifies the magnitude of the residual noise, and detects those fabrication errors that are not common to the device and the undoing procedure; photon loss and dark counts are removed beforehand by heralding and are therefore outside its scope. By testing the validity of the implementation, the test can corroborate the theoretically proven features of the 3D QRNG, such as strong incomputability and unpredictability, or indicate how the implementation should be corrected, if necessary. The test can be easily incorporated into the QRNG and used as a means of experimental certification. The test module is fabricated identically to the QRNG.
\end{abstract}

\section{Introduction}
\label{s:intro}

The algorithms used in cryptography are highly dependent on the quality of the random numbers used during the encryption processes; hence, the need for Quantum Random Number Generators (QRNG) is increasingly understood and 
accepted, especially under the expectation that quantum computers may be able to break encryption methods previously considered secure. Genuine randomness, certified by a mathematical proof of maximal unpredictability, can only be achieved (to date) by Quantum Random Number Generators that measure value indefinite quantum observables, like the 3D-photonic QRNG~\cite{RSPA23}. 
In this context, it is essential to provide a broader range of users with a simple empirical test to certify the 3D QRNGs, not just those who can afford expensive technology and laboratories. 

We use theoretical and experimental certification to guarantee genuine randomness.
Value indefiniteness, which informally means that the quantum random digit is created ``out of nothing'', and not by scrambling pre-existing digits, is the best theoretical certification, and it can only be provided by applications of the Kochen-Specker Theorem, in its localized form~\cite{abbott2012strongrandomness,vi-aeverywhere-2014,2015-AnalyticKS} studied in~\cite{RSPA23}. The Localized Kochen-Specker Theorem identifies value indefinite observables, thereby certifying their randomness. Bell's Theorem offers additional certification based on nonlocality and device-independence~\cite{PironioEtAl2010RandomNumbersCertifiedByBellsTheorem}, but this does not guarantee superiority over other QRNGs based on value-indefiniteness. It is more challenging to build with small hardware components because it requires simultaneous measurements of entangled states at sufficiently large distances to ensure that no local interaction has occurred.

The 3D QRNGs~\cite{qrng2020,RSPA23} are certified, in addition to the Kochen-Specker Theorem, by the following mathematical results.
\begin{itemize}
	\item 
Every quantum random sequence is 3-bi-immune,  by Theorems 6 and 7 in~\cite{qrng2020}. This is stronger than bi-immunity, which is stronger than incomputability; see also~\cite{CaludeEtAl2021BiImunity}.
	\item 
Every quantum random sequence is maximally unpredictable, meaning that no algorithm can accurately predict any of its digits, Theorem 8 in~\cite{qrng2020}.
	\item 
Every quantum random sequence is Borel normal, which means that in base $b$, every block of length $k$ occurs with limiting frequency $b^{-k}$, see Lemma 1 in~\cite{qrng2020}.
\end{itemize}

These results provide strong theoretical evidence of genuine randomness. 
However, the physical realization of any device, especially a quantum one, faces practical obstacles, which result in errors.
The effect of these errors on value indefiniteness was studied in~\cite{TrejoCaludeStoica2026NDPhotonicQRNGsNoisyEnvironment}. For full generality, in~\cite{TrejoCaludeStoica2026NDPhotonicQRNGsNoisyEnvironment} we use completely positive maps (CP-maps) and Kraus operators to account for the openness of the system. We use POVMs for the measurements, combined with applying the Stinespring Dilation Theorem~\cite{Stinespring1955PositiveFunctionsOnC-star-algebras}. This allows us to control the errors due to imperfect MZI beam splitters and phase shifters, forbidden splitting ratio regions, depolarising noise, detector efficiency and dark counts.

In integrated quantum photonics, a major source of errors are photon loss and dark counts, which occur in the couplings and waveguides, depending on the platform. Additionally, there could be fabrication errors associated with the beam splitters and phase shifters, as well as detection inefficiency. Specific experimental tests can detect these errors. For example, to handle photon loss and detection inefficiency, we can use heralding and coincidence counting, a method used in practice to realize a single-photon source~\cite{MosleyEtal2008HeraldedGenerationOfUltrafastSinglePhotonsInPureQuantumStates}.
In the analysis of the QRNGs based on contextuality from~\cite{TrejoCaludeStoica2026NDPhotonicQRNGsNoisyEnvironment}, we proposed photon heralding to remove the outputs that cannot be certified by value indefiniteness, in addition to realizing a single-photon source.
Likewise, in the present proposal we can apply the testing method to the data already filtered by coincidence counting, because this is what we want to test, rather than to the detections we ignore due to uncertainty in their origin via value indefiniteness. Focusing only on outcomes certified by value indefiniteness reduces the test's burden.

The usual methods to test the randomness of RNGs, like those provided by NIST~\cite{NIST-SP800-22,NIST-SP800-22-RevisionDecision}, cannot probe algorithmic properties of randomness such as incomputability and unpredictability. While the tests proposed in~\cite{AbbottEtAl2018ExperimentallyProbingTheAlgorithmicRandomnessAndIncomputabilityOfQuantumRandomness} provide a strong empirical support, they are not easy to use by the typical user of quantum randomness.

The 3D QRNG is based on measuring a qutrit prepared in a particular state, as reviewed in Section~\ref{s:3D-QRNG}.
This article proposes a method to test whether a photonic 3D QRNG operates in accordance with its underlying theory. The test can be easily incorporated into the QRNG and fabricated identically to it.

Section~\ref{s:physical-constraints-implem} contains a brief discussion of physical constraints of implementations of circuits, in particular those used as QRNGs.

Section~\ref{s:LKS} reviews the theoretical foundation of the 3D-photonic QRNG~\cite{qrng2020,RSPA23}, which is provided by the localized variant of the Kochen-Specker Theorem. This result allows us to ensure that the output values arise from the value indefiniteness of the measured observables, thereby certifying their randomness.
In Section~\ref{s:3D-QRNG}, we review the 3D-photonic QRNG~\cite{qrng2020,RSPA23} and the $N$D QRNG from~\cite{ND-QRNG} based on this result.

In Section~\ref{s:modeling-errors} we discuss the modeling of errors, exemplifying with a simple model that facilitates understanding of the proposed inversion test. However, we will model the errors in full generality, accounting for noise, fabrication errors, photon loss, dark counts, etc. For this, we model the process as a completely positive map and propose eliminating photon loss and dark counts by using heralding photons.

In Section~\ref{s:test-by-undo}, we introduce the inversion test proposed in this article. The idea is to combine the QRNG circuit with a gate-by-gate inversion of the same circuit, motivated by the fact that the simplest verifiable transformation is the identity.
This simple idea turns out to be quite versatile, allowing for more general tests that will be briefly discussed in Section~\ref{s:generalizations}.
We conclude with Section~\ref{s:conclusions}.

\section{Physical constraints of implementations of QRNGs}
\label{s:physical-constraints-implem}

The theory behind the 3D QRNG from references~\cite{qrng2020,RSPA23} and the $N$D QRNG from~\cite{ND-QRNG} is rigorously established and is based on value indefiniteness. We review the theoretical apparatus in Section~\ref{s:LKS}. But no matter how promising the theory may be, it is important to fabricate real-world QRNGs as close as possible to the theoretical one, and to certify them.

For this, we make several considerations.

First, a very high precision is needed to ensure the unitarity, fidelity, correctness of output probabilities (e.g. 1/4, 1/2 and 1/4), and to approximate with high accuracy the parameters of the circuit. Although there will always be a small error, this can be mitigated under certain conditions by  using randomness extractors, see~\cite{herrero2017quantum} (von Neumann method is an extractor, but it only works for a very limited type of distributions~\cite{AbbottEtAl2012VonNeumannNormalisationOfAQRNG}).

Second, there may be a problem regarding the separation of photon counts, but the standard solution of using heralding photons allows us to remove the situations when no detector or multiple detectors are activated~\cite{TrejoCaludeStoica2026NDPhotonicQRNGsNoisyEnvironment}.

Third, there are theoretical limitations of the practical realizability of quantum components. An important example is expressed by the Wigner-Araki-Yanase Theorem and its generalizations (WAY), which limit the simultaneous accuracy and repeatability of measurements of observables that do not commute with additively conserved quantities~\cite{Wigner1952MessungQMOperatorenPBusch2010EnTranslation,ArakiYanase1960MeasurementofQMOperators,LoveridgeBusch2011MeasurementofQMOperatorsRevisited}. This applies, in particular, to spin measurements~\cite{Wigner1952MessungQMOperatorenPBusch2010EnTranslation}.
In addition, our recent result~\cite{Stoica2025CanWeAccuratelyReadWriteQuantumData} shows that the limitations of accuracy occur both for reading and writing quantum data. Still, since no lower bound is known, the result from~\cite{Stoica2025CanWeAccuratelyReadWriteQuantumData} is compatible with approaching accuracy as close as desired. The limitations imposed by the WAY Theorem are not restricted to quantum measurements, but also apply to quantum logic gates, imposing strict limitations to quantum computing~\cite{Ozawa2004WAY-QuantumComputer,Ozawa2002ConservativeQuantumComputing}. 
Fortunately, according to Lidar~\cite{Lidar2003CommentOnConservativeQuantumComputing}, even without explicitly taking into consideration the limits imposed by the WAY Theorem, the quantum logic gates are constructed in practice in the right way to avoid these problems, by constraining the states to a subspace of the Hilbert space that is invariant to unitary evolution. Since this relies on the fidelity of the components produced by manufacturers, and since the full scope of the WAY Theorem and its generalizations is still unknown~\cite{LoveridgeBusch2011MeasurementofQMOperatorsRevisited}, one has to rely on empirical verification of the fidelity of the device, and the inversion test proposed here offers a way to do this.

For maximal generality, which includes the above-mentioned sources of error, we model the circuits as completely positive maps, as we did for the theoretical validation of value indefiniteness under error from reference~\cite{TrejoCaludeStoica2026NDPhotonicQRNGsNoisyEnvironment}. 
Let us recall the definition of a completely positive map (CP map).
A \emph{positive map} $\obs{C}$ is a linear map $\obs{C}$ between two $C^\ast$-algebras sending positive elements into positive elements.
In our case, the $C^\ast$-algebras are the algebras of bounded operators on  Hilbert spaces, so for two Hilbert spaces $\hilbert_1$ and $\hilbert_2$, $\obs{C}:\mathcal{B}(\hilbert_1)\to\mathcal{B}(\hilbert_2)$.
A positive map is \emph{completely positive} if its induced maps between the $k\times k$ matrix algebras with entries in the algebras of bounded operators, $\C^{k\times k}\otimes \mathcal{B}(\hilbert_1)\to\C^{k\times k}\otimes \mathcal{B}(\hilbert_2)$, are positive for all $k$.

Ideally, the circuit obtained by combining the QRNG circuit and the mirror circuit, used for the inversion testing, should implement the identity transformation.
Since the identity transformation is much easier to measure, measurements and calibration are easier, regardless of the sources of error.

In addition to the experimental certification that can be obtained as detailed in reference~\cite{AbbottEtAl2018ExperimentallyProbingTheAlgorithmicRandomnessAndIncomputabilityOfQuantumRandomness}, it is important to offer the users the possibility to convince themselves that the random numbers generated are genuinely quantum, rather than being due to noise.
To make sure that the device works and the produced photons have the right properties, we suggest two optional modules:
\begin{enumerate}
	\item 
A module that can be coupled to the 3D QRNG and count the output photons, to verify the given probabilities, e.g. $1/4$, $1/2$, $1/4$. 
	\item 
A module that can undo the unitary transformation, and test whether the initial state of the photon was recovered, which will be discussed in Section~\S\ref{s:test-by-undo}.
\end{enumerate}

In conjunction with statistical and randomness tests like the conservation of 3-bi-immunity~\cite{qrng2020} and the fact that it relies on the Localized Kochen-Specker Theorem~\cite{abbott2012strongrandomness,vi-aeverywhere-2014,2015-AnalyticKS,RSPA23}, they can assure the users that the 3D QRNG is actually quantum, and its outputs satisfy the predictions of the quantum model in~\cite{qrng2020}.

\section{Localized Kochen-Specker Theorem}
\label{s:LKS}

The localized variant of the Kochen-Specker Theorem gives a practical characterization of quantum measurements whose outcomes are indefinite, that is, that are not determined by pre-existing properties of the system.
This can be used to produce genuinely random numbers via quantum means, as in the case of the 3D-photonic QRNG.

Let $\mathbb{C}^n$, $n>2$ be a complex Hilbert space. Let $\mathcal{O} \subseteq \big\{\obs{P}_{\psi}:=\dyad{\psi} \mid \ket{\psi} \in \mathbb{C}^{n},\braket{\psi}{\psi}=1 \big\}$ be a nonempty set of one-dimensional projection observables on $\mathbb{C}^n$.
A \emph{context} is a subset $\mathcal{C} \subset \mathcal{O}$ of $\mathcal{O}$ with $n$ elements so that for all $\obs{P}_{\psi}, \obs{P}_{\phi} \in \mathcal{C}$ with $\obs{P}_{\psi} \neq \obs{P}_{\phi}, \braket{\psi}{\phi}=0$. 

A \emph{value assignment function} on $\mathcal{O}$ is a partial function $v:\mathcal{O}\rightharpoonup\{0,1\}$, with possible indefinite values for some observables in $\mathcal{O}$. The symbol ``$\rightharpoonup$'' represents the fact that $\operatorname{dom}v$, the domain of $v$, is not necessarily the entire set $\mathcal O$.
If $v(\obs{P})\in\{0,1\}$, the observable $\obs{P}$ is \emph{value definite}; if $v(\obs{P})$ is undefined ($P\notin\operatorname{dom}v$), $\obs{P}$ is \emph{value indefinite}. If every observable $\obs{P} \in \mathcal{O}$ is value definite, we say that $\mathcal{O}$ is \emph{value definite}.

Before stating the Localized version of the Kochen-Specker Theorem, we need to define the following conditions.

\begin{itemize}
	\item \textbf{Admissibility condition.}
This condition guarantees agreement with quantum mechanics predictions. Fix a set $\mathcal{O}$ of one-dimensional projection observables in $\mathbb{C}^n$ and the assignment $v:\mathcal{O}\to \{0,1\}$. Then $v$ is admissible if the following two conditions hold for every context $\mathcal{C} \subseteq \mathcal{O}$: a) if there exists $P\in\mathcal{C}$ with $v(P)=1$, then $v(P')=0$, for all $P'\in \mathcal{C} \setminus \{P\}$, b) if there exists $P\in\mathcal{C}$ with $v(P')=0$ for all $P'\in \mathcal{C} \setminus \{P\}$, then $v(P)=1$.
  \item \textbf{Non-contextuality of definite values}. Every outcome obtained by measuring a value-definite observable is \emph{non-contextual}, i.e., it does not depend on other compatible observables which may be measured alongside it.
This condition is expected in classical physics, but it is broken by quantum mechanics, as shown by the Kochen-Specker Theorem~\cite{Kochen:2017aa}.
  \item \textbf{Eigenstate principle.} For a quantum system prepared in the state $\ket{\psi}$, 
the projection observable $\obs{P}_\psi$ is value definite.
\end{itemize}

We are ready to state the Localized Kochen-Specker Theorem.

\begin{theorem}[Localized Kochen-Specker Theorem~\cite{abbott2012strongrandomness,vi-aeverywhere-2014,2015-AnalyticKS}]
\label{thm:LKS}
Assume a quantum system prepared in the state $\ket{\psi}$ in a Hilbert space $\mathbb{C}^n$ with $n\geq 3$, and
let $\ket{\phi}$ be any quantum state such that $0<\abs{ \braket{\psi}{\phi}}<1$.
Let $\mathcal{O}$ be a set of one-dimensional projection observables on $\mathbb{C}^n$ containing $\obs{P}_{\psi}$ and $\obs{P}_{\phi}$, and  $v:\mathcal{O}\to\{0,1\}$ a value assignment function.
If the following three
conditions are satisfied: i) admissibility, ii) non-contextuality, and iii) eigenstate
principle, then the projection observable $\obs{P}_{\phi}$ is value indefinite.
\end{theorem}

In other words, the value obtained by measuring the observed system does not pre-exist, nor do pre-existing data determine it. This is genuine randomness; it is what Wheeler calls an ``elementary act of creation''~\cite{Wheeler1983LawWithoutLaw}.

\section{The 3D-photonic QRNG}
\label{s:3D-QRNG}

The quantum system whose measurement produces quantum random numbers based on Theorem~\ref{thm:LKS} requires a Hilbert space of dimension at least $3$.
We focus on spin-1 observables of a general form
\begin{equation}
\label{eq:spin-one-gen}
\obs{S}(\theta,\varphi)=
\begin{pmatrix}
\cos{\theta} & \frac{e^{-i\varphi}\sin{\theta}}{\sqrt{2}}& 0\\
\frac{e^{i\varphi}\sin{\theta}}{\sqrt{2}} & 0 & \frac{e^{-i\varphi}\sin{\theta}}{\sqrt{2}}\\
0 & \frac{e^{i\varphi}\sin{\theta}}{\sqrt{2}} & -\cos{\theta}
\end{pmatrix}.
\end{equation}

In particular, $\obs{S}_z=\obs{S}(0, 0)$ and $\obs{S}_x=\obs{S}(\pi/2,0)$.

To prepare the system, we first measure the value of the spin-1 observable $\obs{S}_z=\obs{S}(0, 0)$.
To measure the system, we need to choose an operator whose eigenvectors are different but not orthogonal to the prepared state. As in \cite{RSPA23}, we will use a unitary operator $\obs{U}_x$ to implement the measurement of the spin operator $\obs{S}_x=\obs{S}(\pi/2,0)$,
\begin{equation}
\label{eq:Ux}
\obs{U}_x =\frac{1}{2}
\begin{pmatrix}
1 & \sqrt{2} & 1\\
\sqrt{2} & 0 & -\sqrt{2}\\
1 & -\sqrt{2} & 1
\end{pmatrix}.
\end{equation}

In integrated photonics, unitary transformations can be implemented using beam splitters, for example, with \emph{multimode interferometers} (MMIs).
A \emph{Mach-Zehnder interferometer} (MZI) can be obtained by integrating two MMIs with a thermal phase shifter for phase modulation.
Since the performance of MMIs is close to an ideal balanced beam splitter, they can be used to prepare the system.
Therefore, the unitary operator~\eqref{eq:Ux} can be implemented using MZIs constructed from MMIs, with good performance.
In addition to being based on value indefiniteness, this implementation does not use entanglement. It does not require low temperatures, so it is more practical and accessible than, for example, the 3D QRNG from~\cite{PhysRevLett.119.240501}, which requires cooling down to $\sim$20 mK.

The matrix~\eqref{eq:Ux} can be realized in terms of beam splitters and single-mode phase shifts~\cite{reck1994experimental,ClementsEtAl2016OptimalDesignForUniversalMultiportInterferometers,Cilluffo2024ClementsUnitaryDecompositionInPractice} as a product of matrices, each of which, when restricted to some two-dimensional subspace, is of the form
\begin{equation}
\label{eq:matrices-sub}
\obs{T}(\theta, \varphi)
=
\begin{pmatrix}
\cos\theta & i e^{i\varphi} \sin\theta \\
i\sin\theta & e^{i\varphi} \cos\theta \\
\end{pmatrix},
\end{equation}
and the restriction to the orthogonal complement, which is a one-dimensional subspace, is the identity.
Here, $\abs{\cos\theta}^2$ represents the reflectivity coefficient, $\abs{\sin\theta}^2$ the transmittance coefficient, and $\varphi$ represents the phase of an external phase shifter on the second input port.

The unitary matrix $\obs{U}$ to be realized is $N\times N$, where $N$ can be greater than $2$. The unitary $\obs{U}$  is realized using $N\times N$ matrices that act as equation~\eqref{eq:matrices-sub} on the two-dimensional subspace spanned by modes $j,j+1$, and as the identity on its orthogonal complement~\cite{reck1994experimental,ClementsEtAl2016OptimalDesignForUniversalMultiportInterferometers,Cilluffo2024ClementsUnitaryDecompositionInPractice}:
\colorlet{clrCell}{cyan!15}
\begin{equation}
\label{eq:matrices-sub-jk}
\begin{aligned}
\obs{T}_{j,j+1}(\theta, \varphi)
&=
\begin{+pmatrix}[cell{1-3}{1-3}={clrCell}]
\obs{I}_{j-1} & 0 & 0 \\
0 & \obs{T}(\theta, \varphi) & 0 \\
0 & 0 & \obs{I}_{N-j-1} \\
\end{+pmatrix}\\
&=
\begin{+pmatrix}[cell{1-3}{1-3}={clrCell},cell{4,5}{4,5}={clrCell},cell{6-8}{6-8}={clrCell}]
1 & \cdots & 0 & 0 & 0 & 0 & \cdots & 0 \\ 
\vdots & \ddots & \vdots & \vdots & \vdots & \vdots & \ddots & \vdots \\
0 & \cdots & 1 & 0 & 0 & 0 & \cdots & 0 \\
0 & \cdots & 0 & \cos\theta & i e^{i\varphi} \sin\theta & 0 & \cdots & 0 \\
0 & \cdots & 0 & i\sin\theta & e^{i\varphi} \cos\theta & 0 & \cdots & 0 \\
0 & \cdots & 0 & 0 & 0 & 1 & \cdots & 0 \\
\vdots & \ddots & \vdots & \vdots & \vdots & \vdots & \ddots & \vdots \\
0 & \cdots & 0 & 0 & 0 & 0 & \cdots & 1 \\ 
\end{+pmatrix}
\end{aligned}
\end{equation}
where $\obs{I}_k$ is the $k \times k$ identity matrix.

Then, any $N \times N$ unitary matrix $\obs{U}$ can be written in the form~\cite{ClementsEtAl2016OptimalDesignForUniversalMultiportInterferometers}
\begin{equation}
\label{eq:clements}
\obs{U}=\obs{D}\cdot \obs{T}_{j_1,j_1+1}(\theta_1, \varphi_1)\cdot \obs{T}_{j_2,j_2+1}(\theta_2, \varphi_2)\cdot \ldots \cdot \obs{T}_{j_k,j_k+1}(\theta_k, \varphi_k),
\end{equation}
where the diagonal matrix $\obs{D}$ has as diagonal entries only complex numbers of the form $e^{i\phi}$.
In our case, the matrix~\eqref{eq:Ux} can be decomposed as a product involving such matrices in the following ways~\cite{RSPA23}:
\begin{equation}
\label{eq:Ux-decomposed}
\obs{U}_x = \obs{B}_{1,2}^{-1}\cdot \obs{B}_{2,3} \cdot \obs{D} \cdot \obs{B}_{1,2}= \obs{D}' \cdot \obs{B}_{1,2}'\cdot \obs{B}_{2,3} \cdot \obs{B}_{1,2},
\end{equation}
where the matrices representing the beam splitters are 
\begin{equation}
\label{eq:matrices-beam-splitters}
\begin{aligned}
\obs{B}_{1,2} &=
\begin{pmatrix}
\sqrt{\frac{2}{3}} & \frac{1}{\sqrt{3}} & 0\\
\frac{i}{\sqrt{3}} & -i\sqrt{\frac{2}{3}} & 0\\
0 & 0 & 1
\end{pmatrix},
&
\obs{B}_{2,3} &=
\begin{pmatrix}
1 & 0 & 0\\
0 & \frac{1}{2} & -\frac{i\sqrt{3}}{2}\\
0 & \frac{i\sqrt{3}}{2} & -\frac{1}{2}
\end{pmatrix}, \\
\obs{B}_{1,2}^{-1} &=
\begin{pmatrix}
\sqrt{\frac{2}{3}} & -\frac{i}{\sqrt{3}} & 0\\
\frac{1}{\sqrt{3}} & i\sqrt{\frac{2}{3}} & 0\\
0 & 0 & 1
\end{pmatrix},
&
\obs{B}_{1,2}' &=
\begin{pmatrix}
\sqrt{\frac{2}{3}} & \frac{i}{\sqrt{3}} & 0\\
-\frac{i}{\sqrt{3}} & -\sqrt{\frac{2}{3}} & 0\\
0 & 0 & 1
\end{pmatrix}.
\end{aligned}
\end{equation}

The matrices representing the single-mode phase-shifts are
\begin{equation}
\label{eq:matrices-phase-shifters}
\begin{aligned}
\obs{D}&=\begin{pmatrix} 
1 & 0 & 0\\
0 & -1 & 0\\
0 & 0 & -1
\end{pmatrix},
&
\obs{D}'&=\begin{pmatrix} 
1 & 0 & 0\\
0 & i & 0\\
0 & 0 & -1
\end{pmatrix}.
\end{aligned}
\end{equation}

The parameters $\theta$ and $\varphi$ from equation~\eqref{eq:matrices-sub}, for the matrices~\eqref{eq:matrices-beam-splitters}, are represented in Table~\ref{table:Bmn-matrices}.

\begin{table}[!ht]
\centering
\begin{tabular}{|l|r|r|}
     \hline
$\obs{B}_{m,n}$ & $\theta$ & $\varphi$ \\
     \hline
$\obs{B}'_{1,2}$ & $-\acos\sqrt{\frac{2}{3}}$ & $\pi$ \\
$\obs{B}_{2,3}$ & $\frac{2\pi}{3}$ & $\pi$ \\
$\obs{B}_{1,2}$ & $\acos\sqrt{\frac{2}{3}}$ & $-\frac{\pi}{2}$ \\
\hline
\end{tabular}
\caption{Angles and phases for the beam splitter operators.}
\label{table:Bmn-matrices}
\end{table}

The physical realization of the universal unitary decomposition $\obs{U}_x$ by means of the three-mode multiport interferometer is represented in Figure~\ref{3D-QRNG}, which is reproduced from~\cite{RSPA23}.

\figpdf{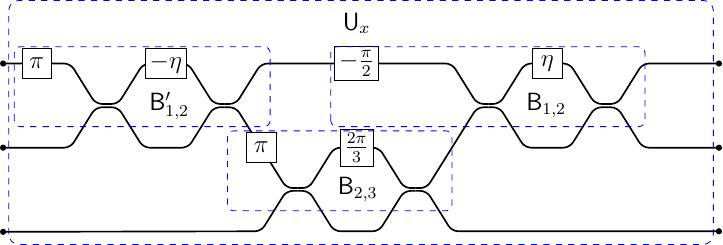}{Reproduced from~\cite{RSPA23}. Physical realization of the universal unitary decomposition $\obs{U}_x$ by means of a three-mode multiport interferometer. An arrangement of Mach–Zehnder interferometers consisting of phase shifters and balanced directional couplers illustrates its construction. Here, $\eta=\acos\frac{\sqrt{2}}{3}$.}{!ht}

\section{Modeling errors}
\label{s:modeling-errors}

The decomposition of $\obs{U}_x$ from~\eqref{eq:Ux-decomposed} is an idealization; in practice, there will be errors.
A basic model of the errors is to allow the parameters $\theta$ and $\varphi$ to be slightly different from those in Table~\ref{table:Bmn-matrices}, for each of the matrices in equation~\eqref{eq:matrices-beam-splitters}, as in Table~\ref{table:Bmn-matrices-err}.

\begin{table}[!ht]
\centering
\def\arraystretch{1.35}
\begin{tabular}{|l|r|r|}
     \hline
$\widetilde{\obs{B}}_{m,n}$ & $\theta$ & $\varphi$ \\
     \hline
$\widetilde{\obs{B}}'_{1,2}$ & $\theta_1$ & $\varphi_1$ \\
$\widetilde{\obs{B}}_{2,3}$ & $\theta_2$ & $\varphi_2$ \\
$\widetilde{\obs{B}}_{1,2}$ & $\theta_3$ & $\varphi_3$ \\
     \hline
\end{tabular}
\caption{Inaccurate angles and phases for the beam splitter operators.}
\label{table:Bmn-matrices-err}
\end{table}

But this restricted model of coherent fabrication or calibration offsets is by far insufficient, since a realistic model must account for generic open-system noise, stochastic phase noise, imperfect MZI transfer matrices, decoherence, photon loss, and detection inefficiency.
Reference~\cite{TrejoCaludeStoica2026NDPhotonicQRNGsNoisyEnvironment} takes these sources of noise and unsharpness into account by treating the system as a completely positive map with Kraus operators and a depolarizing qutrit or qudit channel, and includes a discussion of detector efficiency and dark counts. We will apply the same method here to both the device to be tested and the mirror device.
Let us first describe the basic model, as an example.

Following~\cite{TrejoCaludeStoica2026NDPhotonicQRNGsNoisyEnvironment}, the matrices representing the beam splitters are:
\begin{equation}
\label{eq:matrices-beam-splitters-err}
\begin{aligned}
\widetilde{\obs{B}}_{1,2} &=
\begin{pmatrix}
\obs{T}(\theta_3,\varphi_3) & 0\\
0 & 1
\end{pmatrix},
&
\widetilde{\obs{B}}_{2,3} &=
\begin{pmatrix}
1 & 0 \\
0 & \obs{T}(\theta_2,\varphi_2) \\
\end{pmatrix}, \\
\widetilde{\obs{B}}_{1,2}^{-1} &=
\begin{pmatrix}
\obs{T}^\dagger(\theta_3,\varphi_3) & 0\\
0 & 1
\end{pmatrix},&
\widetilde{\obs{B}}_{1,2}' &=
\begin{pmatrix}
\obs{T}(\theta_1,\varphi_1) & 0\\
0 & 1
\end{pmatrix},
\end{aligned}
\end{equation}
and the matrices representing the single-mode phase-shifts are
\begin{equation}
\label{eq:matrices-phase-shifters-err}
\begin{aligned}
\widetilde{\obs{D}}&=\begin{pmatrix} 
e^{i\phi_1} & 0 & 0\\
0 & e^{i\phi_2} & 0\\
0 & 0 & e^{i\phi_3}
\end{pmatrix},
&
\widetilde{\obs{D}}'&=\begin{pmatrix} 
e^{i\phi_1'} & 0 & 0\\
0 & e^{i\phi_2'} & 0\\
0 & 0 & e^{i\phi_3'}
\end{pmatrix},
\end{aligned}
\end{equation}
where $\phi_1\approx\phi_1'\approx 0$, $\phi_2\approx\phi_3\approx\phi_3'\approx\pi$, and $\phi_2'\approx\pi/2$.

Then, instead of equation \eqref{eq:Ux-decomposed} we have
\begin{equation}
\label{eq:Ux-decomposed-err}
\widetilde{\obs{U}}_x = \widetilde{\obs{B}}_{1,2}^{-1}\cdot \widetilde{\obs{B}}_{2,3} \cdot \widetilde{\obs{D}} \cdot \widetilde{\obs{B}}_{1,2}= \widetilde{\obs{D}}' \cdot \widetilde{\obs{B}}_{1,2}'\cdot \widetilde{\obs{B}}_{2,3} \cdot \widetilde{\obs{B}}_{1,2}.
\end{equation}

The imperfect 3D QRNG is represented in Figure~\ref{3D-QRNG-err}.

\figpdf{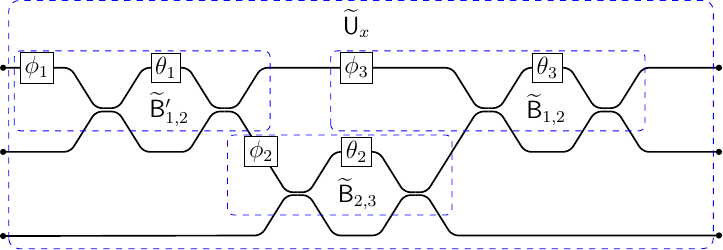}{Imprecise 3D-QRNG.}{!ht}

The extra MMI phases are represented by additional diagonal phase matrices in equation~\eqref{eq:Ux-decomposed-err}.

This basic model still assumes that the process is unitary, but the noise and decoherence require treating the state as a positive operator-valued measure (POVM), and the process as a completely positive map rather than a unitary transformation. Treating the process as a different unitary process would limit the testing method proposed here to verifying that the fabrication errors are consistent. Still, it would ignore the more general and more realistic errors.
To account for imperfections, as a first generalization of the basic model from above, in~\cite{TrejoCaludeStoica2026NDPhotonicQRNGsNoisyEnvironment} we assumed imperfect beam-splitters with a range of losses based on the fidelity of actual components. Then we showed that, under certain conditions, the process remains within the scope of the Kochen-Specker Theorem, guaranteeing maximum unpredictability. Then we generalized it to unsharp measurements (POVMs), by applying the Localized Kochen-Specker Theorem to the pointer states, under the assumption that pointer states are orthogonal, being macroscopic.
For this article, all we need is to generalize the unitary $\widetilde{\obs{U}}_x$ from equation~\eqref{eq:Ux-decomposed-err} to a completely positive map $\widetilde{\obs{C}}_x:\mathcal{B}(\C^3)\to\mathcal{B}(\C^3)$, which in the unitary case is of the form $\widetilde{\obs{C}}_x[\obs{A}]=\widetilde{\obs{U}}_x\obs{A}\widetilde{\obs{U}}_x^\dagger$.

We write $\Cxt$ in Kraus form with a distinguished leading term,
\begin{equation}
    \Cxt[\rho]\;=\; (1-p)\, \Uxt \rho \Uxt\dg \;+\; p \sum_{k \geq 1} \mathsf{K}_k \rho \mathsf{K}_k\dg, \qquad \sum_{k\geq 1}\mathsf{K}_k\dg \mathsf{K}_k = I_3.
\end{equation}
The \emph{coherent} parameters are then the angles and phases of Table 2, which enter through $\Uxt$; we collect their deviations from Table 1 as follows.

\begin{equation}
    \varepsilon \;=\; \big\| \log\big(\Ux\dg \Uxt \big) \big|_2 \;\leq\; \sum_{k+1}^{3}\big( |\Delta\theta_k| + |\Delta\varphi_k| \big) +\sum_{m=1}^3|\Delta|\phi_m|.
\end{equation}

Dark counts and photon loss do not appear here since heralding and coincidence counting handle it before the test applications, and $p$ collects incoherent events not of the form of stochastic phase noise, decoherence, mode crosstalk, and residual detector effects. The worst-case bound is then 
\begin{equation}
    p_{\max}'\;\leq\; (1-p)\, p_{\max} + p,
\end{equation}
with $p_{max}$ the largest ideal outcome probability. (e.g. 1/2 for a 1/4,1/2,1/4 distribution).

Since a CP map is not invertible unless it is unitary, and noting that if $p>0$ there is no completely positive $\Phi$ with $\Phi\circ\Cxt = \mathrm{id}$; then the undoing apparatus implements a second, independently noisy channel $\Cxtm$  of the same form, allowing us to make the mechanism quantitative. Moreover, in the device's composition with the undoing apparatus, the leading Kraus weights multiply, yielding the weight $(1-p)^2$. That is, the coherent and incoherent errors compose oppositely, with coherent deviations partly canceling.

\section{Testing unitarity by undoing the QRNG}
\label{s:test-by-undo}

The test aims
to verify the unitarity of the realized QRNG, and
to assess the accuracy of its physical implementation by checking that the unpredictable errors, and the systematic errors (\textit{i.e.}~consistent deviations in measurements due to imperfect equipment or flawed experimental setups), are sufficiently small.

First, we explain how to construct a module that undoes the unitary transformation and recovers the photon's initial state. To illustrate the idea, we first describe how it can be done for a simple phase shift gate $\obs{P}(\varphi)$, which maps $\ket{0}\mapsto\ket{0}$ and $\ket{1}\mapsto e^{i\varphi}\ket{1}$, as in Figure~\ref{fig-PS}.

\figpdf{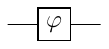}{Phase shift gate.}{!ht}

To test this by reconstructing the original state, we add another phase shift gate with opposite phase, as in Figure~\ref{fig-PS-inv}.

\figpdf{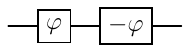}{Undoing a phase shift gate.}{!ht}

This will recover the original state.
Since the unitary matrix representing the phase shift is
\begin{equation}
\label{eq:PS}
\obs{P}(\varphi)
=
\begin{pmatrix}
1 & 0 \\
0 & e^{i\varphi} \\
\end{pmatrix},
\end{equation}
to undo was to multiply by its inverse, which is $\obs{P}(\varphi)^{-1}=\obs{P}(\varphi)^\dagger=\obs{P}(-\varphi)$,
\begin{equation}
\label{eq:PS-inv}
\obs{P}(\varphi)\obs{P}(\varphi)^{-1}
=
\begin{pmatrix}
1 & 0 \\
0 & e^{i\varphi} \\
\end{pmatrix}
\begin{pmatrix}
1 & 0 \\
0 & e^{-i\varphi} \\
\end{pmatrix}
=\begin{pmatrix}
1 & 0 \\
0 & 1 \\
\end{pmatrix}.
\end{equation}

The same procedure can be applied to the $3\times 3$ beam splitter 3D QRNG: take another $3\times 3$ circuit, configured to implement the inverse unitary operation, and connect it to the one that we want to test.

The main version of the inversion test is intended to verify whether the implementation is sufficiently close to a unitary.
One can first test each gate using its inverse, and after that, the full circuit.
The unitary decomposition into gates is non-unique. Such a test is easier if the inverse unitary operation is implemented by inverting the gates used in the direct implementation, in reverse order. 
For example, if the factorization is $U=U_1\cdots U_k$, the inverse has to implement $U^\dagger=U_k^\dagger\cdots U_1^\dagger$, gate by gate. 
Physically, different decompositions will have different properties, resulting in different depth, accumulated loss, component count, and sensitivity to imperfect splitting ratios, as seen in the section on the effect of experimental imperfections in reference~\cite{TrejoCaludeStoica2026NDPhotonicQRNGsNoisyEnvironment}. For example, the Reck and Clements decompositions are different.
Inverting the circuit gate by gate allows the identification of sources of error and correcting them.
Other implementations of the inverse unitary operation of a particular decomposition are also possible, but this would make it more difficult to identify the sources of error, due to the asymmetry.

However, using a mirror device implemented differently from the QRNG may also be useful. Consider a scenario in which Alice builds a black-box QRNG, giving access only to the inputs and the outputs.
Then, a user Bob can use an inverse device to test Alice's device.
This allows Bob to treat the device as an untrusted device, closer to device-independent certification~\cite{PironioEtAl2010RandomNumbersCertifiedByBellsTheorem}. In this use case of the inversion tester, Bob does not calibrate Alice's device, it only verifies that it implements the correct transformation. Bob can report back to Alice and ask for a better calibration, or he can simply use the inversion test as an easy way to verify that the transformation is approximately the intended one.
But this article focuses only on the first application, to quantify the fabrication errors.

The reason why this method recovers the original state, if the implementation is ideal, is that the second device performs the inverse unitary transformation performed by the first device. Then, the output can be tested to verify that it matches the input and to measure the error.
To verify the identity, the test should be done by sending photons into each of the inputs, and verify that the corresponding detector detects them.

Next, we consider an imperfect implementation of the unitary transformation from equation~\eqref{eq:Ux-decomposed-err}.
We assume that all the matrices in this decomposition are CP maps. Ideally, they are unitaries, but errors force us to consider them more general CP maps.
For simplicity, we illustrate the idea using unitaries whose parameters still need to be adjusted, as in equations~\eqref{eq:matrices-beam-splitters-err} and~\eqref{eq:matrices-phase-shifters-err}. Then,
\begin{equation}
\label{eq:Ux-decomposed-err-undo}
\widetilde{\obs{U}}_x\widetilde{\obs{U}}^\dagger_x = \obs{D}' \cdot \widetilde{\obs{B}}_{1,2}'\cdot \widetilde{\obs{B}}_{2,3} \cdot \widetilde{\obs{B}}_{1,2}
\cdot
\widetilde{\obs{B}}_{1,2}^\dagger \cdot \widetilde{\obs{B}}_{2,3}^\dagger \cdot \widetilde{\obs{B}}_{1,2}^{'\dagger} \cdot \obs{D}^{'\dagger}=\obs{I}_3.
\end{equation}

This means that, to test the imperfect QRNG from Figure~\ref{3D-QRNG-err}, we can use its mirror image to undo the unitary transformation. Looking at equation~\eqref{eq:matrices-sub}, we notice that
\begin{equation}
\label{eq:matrices-sub-struct}
\begin{aligned}
\obs{T}(\theta, \varphi)^\dagger
&=
\(\obs{T}(\theta, 0)\obs{T}(0, \varphi)\)^\dagger \\
&=
\begin{pmatrix}
1 & 0 \\
0 & e^{i\varphi}
\end{pmatrix}^\dagger
\begin{pmatrix}
\cos\theta & i \sin\theta \\
i\sin\theta & \cos\theta \\
\end{pmatrix}^\dagger \\
&=
\begin{pmatrix}
1 & 0 \\
0 & e^{-i\varphi}
\end{pmatrix}
\begin{pmatrix}
\cos(-\theta) & i \sin(-\theta) \\
i\sin(-\theta) & \cos(-\theta) \\
\end{pmatrix} \\
&=\obs{T}(0, -\varphi)\obs{T}(-\theta, 0).\\
\end{aligned}
\end{equation}

In general, for a unitary matrix $\obs{U} = \obs{T}_{j_1,j_1+1}(\theta_1, \varphi_1) \ldots \obs{T}_{j_k,j_k+1}(\theta_k, \varphi_k)$ as in equation~\eqref{eq:clements},
\begin{equation}
\label{eq:clements-inverted}
\begin{array}{rcl}
\obs{U}^\dagger
&=&\(\obs{T}_{j_1,j_1+1}(\theta_1, \varphi_1) \ldots \obs{T}_{j_k,j_k+1}(\theta_k, \varphi_k)\)^\dagger\\
&=&\obs{T}_{j_k,j_k+1}(\theta_k, \varphi_k)^\dagger \ldots  \obs{T}_{j_1,j_1+1}(\theta_1, \varphi_1)^\dagger\\
&=&\obs{T}_{j_k,j_k+1}(0, -\varphi_k)\obs{T}_{j_k,j_k+1}(-\theta_k, 0) \ldots \\
&&\ldots \obs{T}_{j_1,j_1+1}(0, -\varphi_1)\obs{T}_{j_1,j_1+1}(-\theta_1, 0).\\
\end{array}
\end{equation}

\begin{figure}[!ht]
    \centering
        \includegraphics[width=\textwidth]{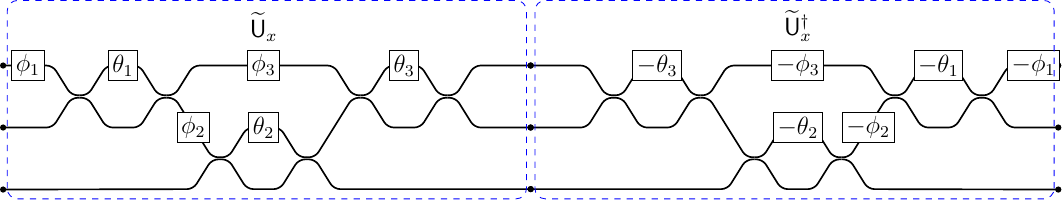}
    \caption{\small{Testing the 3D-QRNG by undoing. At the outputs of the 3D-QRNG implementing the transformation $\widetilde{\obs{U}}_x$, we connect the inputs of a mirrored version, which implements the inverse transformation $\widetilde{\obs{U}}_x^\dagger$, and verify that we recover the input state. We also make sure to compensate for the path differences, since they can introduce phase shifts.}}
    \label{3D-QRNG-B2B-err}
    \end{figure}

From these observations, we see that to obtain the mirror QRNG, the order of the components has to be reversed, and the angles and phases must be negated, as in Figure~\ref{3D-QRNG-B2B-err}.

For the most general case, we use a CP map $\widetilde{\obs{C}}_x$ instead of a unitary $\widetilde{\obs{U}}_x$.
Equation~\eqref{eq:Ux-decomposed-err-undo} holds only approximately
\begin{equation}
\label{eq:Cx-decomposed-err-undo}
\Cxtm \circ \Cxt \;\approx\; \mathrm{id}_{\mathcal{B}(\mathbb{C}^3)},
\end{equation}
where $\mathrm{id}_{\mathrm{B}(\mathbb{C}^3)}$ is the identity map on $\mathcal{B}(\C^3)$. Expanding both channels, the identity branch carries weight $(1-p)^2$, so for the incoherent part, the deviation from the identity map is $2p+O(p^2)$ plus a coherent part discussed next.

In practice, we can only verify the probability that the output corresponds to the input for each input. More precisely, we sent photons through each mode and counted the ratios detected by the corresponding output detector.
This allows us to verify the CP map on the projectors $\dyad{j}$. 
After a large number of runs, we obtain a $3\times 3$ matrix $V$ containing the probabilities for the transitions $j\mapsto k$, for $j,k\in\{1,2,3\}$. We then compare it with the identity matrix.
Briefly, the result of the test is quantified in terms of the probability that each output of the mirror QRNG matches the corresponding input of the direct QRNG. The probabilities are defined as the ratio of counts for each input-output pair, for a large number of repetitions. They are used to construct a stochastic matrix and compare it with the identity matrix.

To make the comparison quantitative, we define the residual
\begin{equation}
    R\;=\; 1 - \tfrac13 \operatorname{tr} V,
\end{equation}
so that $R=0$ for a perfect test.
Now, $R$ responds differently to coherent and incoherent errors. By (\ref{eq:Cx-decomposed-err-undo}), we have $R\simeq2p$ for the incoherent case, while for coherent errors, due to squared amplitudes in $V$ we have $R=\Theta(\varepsilon^2)$. Nonetheless, considering coherent errors where the fabrication offset is an odd function of the design parameters, $R$ vanishes. Thus this applies to errors not of this form such as constant angular offset common to all elements.

If the resulting matrix is sufficiently close to the identity, this will show that
\begin{enumerate}
	\item 
even if there are errors, they do not break the unitarity of the device,
	\item 
if the QRNG device and its inverted version are fabricated identically (but with opposite phase shifters in reverse order, as described), then their systematic errors cancel in the composition, and the test bounds only the part of the error that differs between the two circuits. Thus, we can quantify and correct.
\end{enumerate}

A schematic representation is given in Figure~\ref{fig-Cx-inv}.

\figpdf{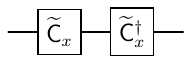}{Undoing the completely positive map $\widetilde{\obs{C}}_x$.}{!ht}

Then, given that the resulting matrix contains squared amplitudes, we want to verify that the phases are undone too.
One way to test this is to compare the output after inversion with the input using interference.
More precisely, use a beam splitter to split a photon into two equal-amplitude components, pass a component through the implementation of $\Cxtm\circ\Cxt$ and after that, instead of detecting the output photons, we make them interfere with the other components. To verify that the phase differences are minimal, we check if the interference is constructive.

Since the procedure returns the pair $(\epsilon,p)$ we can use the perturbation bound $|\sqrt{p'_i} - \sqrt{p_i}|\leq\epsilon$ together with $(14)$ to obtain a min-entropy estimate per digit
\begin{equation}
    \Hmin \;\geq\; -\log_2\!\Big[(1-p)\big(\sqrt{p_{\max}} + \varepsilon\big)^2 + p \Big].
\end{equation}

\begin{remark}
\label{rem:allow}
Note that the Localized Kochen-Specker Theorem was shown to be very general~\cite{2015-AnalyticKS}: to ensure value indefiniteness, it is necessary and sufficient that $0<\abs{ \braket{\psi}{\phi}}<1$, as stated in Theorem~\ref{thm:LKS}. Here, $\ket{\psi}$ is the prepared state, and $\dyad{\phi}$ is the projector corresponding to the activated detector.
For a $3\times 3$ circuit, all detectors that satisfy this condition can be used for random numbers, as they originate in value indefiniteness.
Therefore, we do not need to ensure that the operator $\widetilde{\obs{U}}_x$ is exactly the operator $\obs{U}_x$ from equation \eqref{eq:Ux}: it is sufficient to ensure that it realizes a sharp measurement and that it satisfies the conditions of Theorem~\ref{thm:LKS}.
That is, we will allow systematic errors 
as long as their effect is kept under control and can be undone by additional phase shifts to test that all conditions are met.
A positive unitarity test will show that orthogonal vectors in the input are mapped to orthogonal vectors in the output, ensuring sharpness.
If every output has a nonzero detection probability, even when the input is in only one mode, the condition $0<\abs{\braket{\psi}{\phi}}<1$ is also satisfied by all one-photon detector events. More precisely, for a prepared state $\ket{\psi}$ and a detector corresponding to $\dyad{\phi}$, $0<\abs{\braket{\psi}{\phi}}$ is satisfied if that detector is triggered after sufficiently many runs, and $\abs{\braket{\psi}{\phi}}<1$ is satisfied if at least another detector is triggered after sufficiently many runs.
Therefore, based on Theorem~\ref{thm:LKS}, if the inequality $0<\abs{ \braket{\psi}{\phi}}<1$ is verified, passing the test of unitarity by inversion should be sufficient to ensure the value indefiniteness of the generated digits.
\end{remark}

\section{Generalizations}
\label{s:generalizations}

The inversion-based test can be applied to implementations of other unitary matrices, including $N\times N$ matrices with $N>3$. In particular, it works for testing the $N$-dimensional QRNG proposed in~\cite{ND-QRNG}.

In addition, it is possible to perform tests by coupling more imperfect copies of the QRNG and their mirror images.
This will amplify some of the potential errors and make their detection and quantification easier.

A possible arrangement consists of placing all the mirror images at the end of the circuit
\begin{equation}
\label{eq:test-multiple-aabb}
\left(\widetilde{\obs{U}}_{x,1}\cdot\ldots\cdot\widetilde{\obs{U}}_{x,n}\right)\cdot\left(\widetilde{\obs{U}}_{x,n}^{'\dagger} \cdot\ldots\cdot\widetilde{\obs{U}}_{x,1}^{'\dagger}\right) \cong\obs{I},
\end{equation}
or by alternating them,
\begin{equation}
\label{eq:test-multiple-abab}
\left(\widetilde{\obs{U}}_{x,1}\cdot\widetilde{\obs{U}}_{x,1}^{'\dagger}\right)\cdot\ldots\cdot\left(\widetilde{\obs{U}}_{x,n}\cdot\widetilde{\obs{U}}_{x,n}^{'\dagger}\right)\cong\obs{I}.
\end{equation}

It is even possible to test any permutation of $n$ QRNGs $\widetilde{\obs{U}}_{x,j}$ and $n$ mirror QRNGs $\widetilde{\obs{U}}_{x,j}^{'\dagger}$, where $j\in\{1,\ldots,n\}$.
This will allow errors to propagate from one mode to another and interfere, revealing phase errors that otherwise could be missed.

Since the unitary from equation~\eqref{eq:Ux} satisfies $\obs{U}_x^\dagger=\obs{U}_x$, in this case, another test is possible: if the implementation is expected to be accurate enough and self-adjoint, we should be able to do the inversion test simply by connecting two copies of the device in Figure~\ref{3D-QRNG-B2B-err}.
The tests that amplify the errors, as in equations \eqref{eq:test-multiple-aabb} and \eqref{eq:test-multiple-abab}, are realized, in this case, simply by connecting an even number of copies of the QRNG,
\begin{equation}
\label{eq:test-multiple-aa}
\widetilde{\obs{U}}_{x,1}\cdot\ldots\cdot\widetilde{\obs{U}}_{x,2n} \cong\obs{I}.
\end{equation}

In this case, the same test will provide not only a complementary verification of the unitarity, but also of the self-adjointness of the operator $\obs{U}_x$.

\section{Conclusions}
\label{s:conclusions}

In summary, it is important to test all aspects of the 3D QRNG, both to ensure the device's quality and to satisfy users. In particular:
\begin{enumerate}
	\item 
the physical testing of possible errors and deviations from the ideal ratios,
	\item 
testing the conservation of 3-bi-immunity,
	\item 
testing the unitarity by using a second, reversed 3D QRNG module, combined with the interference test described in Section~\ref{s:test-by-undo}.
\end{enumerate}

The inversion-based test can be applied to QRNGs of any dimension, including 2D QRNGs based on beam splitters.
It can be applied to any QRNGs that rely on a measurement by a known unitary transformation applied to known prepared state. The test does not apply to QRNGs based on noise or decay, because these processes are irreversible for accessible operations.

The inversion-based test allows us to verify that the process is unitary, but it does not guarantee that it is the one inequation~\eqref{eq:Ux}. However, as explained in Remark~\ref{rem:allow}, this is not necessary: it suffices that a sharp measurement is obtained, and that the conditions of Theorem~\ref{thm:LKS} are satisfied, particularly the inequality $0<\abs{ \braket{\psi}{\phi}}<1$. Post-processing can rectify the resulting probabilities in case they are not the expected ones.

\subsection*{Acknowledgement}

We are indebted to Prof. S. Ma and Prof. M. Zimand and the anonymous meticulous reviewer for their comments, which improved the presentation.

\subsection*{Author Contributions Statement}

The authors JMAT, CSC, and OCS are listed in alphabetical order: each author has contributed to the model, photonic implementation, and generalizations.

\subsection*{Research Funding}

This research did not receive funding.

\subsection*{Data Availability Statement}

All data generated or analyzed during this study are included in this published article.

\bibliographystyle{abbrv}
\bibliography{references}

\end{document}